%% file: 0_bare_jrnl_compsoc.tex
\documentclass[10pt,journal,compsoc]{IEEEtran}
\usepackage{graphicx}

\usepackage{subcaption}
\renewcommand{\arraystretch}{1.5}
\usepackage{makecell}
\usepackage{comment}
\usepackage{booktabs} 
\usepackage{multirow}
\usepackage{array}
\usepackage{amsmath} 
\usepackage{times}
\usepackage{soul}
\usepackage{url}

\usepackage{siunitx} %
\usepackage{makecell, caption, chngcntr} %
\usepackage{booktabs}
\usepackage{algorithm}
\usepackage{algpseudocode}
\usepackage{color}
\usepackage{comment}
\usepackage{amsthm}
\usepackage{amssymb}
\usepackage{pifont}
\newcommand{\cmark}{\ding{51}}%
\newcommand{\xmark}{\ding{55}}%

\usepackage[pagebackref=true,breaklinks=true,colorlinks,bookmarks=false]{hyperref}
\ifCLASSOPTIONcompsoc
  \usepackage[nocompress]{cite}
\else
  \usepackage{cite}
\fi
\ifCLASSINFOpdf
\else
\fi
\begin{document}
%
\title{An Adaptive Black-box Backdoor Detection Method for Deep Neural Networks}
%
%
%
%

\author{Xinqiao~Zhang,~\IEEEmembership{Student Member,~IEEE,}
        Huili~Chen,
        Ke~Huang,~\IEEEmembership{Member,~IEEE,}
        and~Farinaz~Koushanfar,~\IEEEmembership{Fellow,~IEEE}
\IEEEcompsocitemizethanks{\IEEEcompsocthanksitem Xinqiao is with the Department of Electrical and Computer Engineering, UC San Diego, San Diego, CA, 92093 and the Department of Electrical and Computer Engineering, San Diego State University, San Diego, CA 92182.\protect\\
E-mail: x5zhang@eng.ucsd.edu}}
\IEEEtitleabstractindextext{%
\begin{abstract}
With the surge of Machine Learning (ML), An emerging amount of intelligent applications have been developed. Deep Neural Networks (DNNs) have demonstrated unprecedented performance across various fields such as medical diagnosis and autonomous driving. 
While DNNs are widely employed in security-sensitive fields, they are identified to be vulnerable to Neural Trojan (NT) attacks that are controlled and activated by stealthy triggers. 
In this paper, we target to design a robust and adaptive Trojan detection scheme that inspects whether a pre-trained model has been Trojaned before its deployment. 
Prior works are oblivious of the intrinsic property of trigger distribution and try to reconstruct the trigger pattern using simple heuristics, i.e., stimulating the given model to incorrect outputs. As a result, their detection time and effectiveness are limited.  
We leverage the observation that the pixel trigger typically features spatial dependency and propose 
the first trigger approximation based black-box Trojan detection framework that enables fast and scalable search of the trigger in the input space. 
Furthermore, our approach can also detect Trojans embedded in the feature space where certain filter transformations are used to activate the Trojan.
We perform extensive experiments to investigate the performance of 
our approach across various datasets and ML models. 
Empirical results show that our approach achieves a ROC-AUC score of $0.93$ on the public TrojAI dataset~\footnote{https://pages.nist.gov/trojai/docs/data.html}. Our code can be found in https://github.com/xinqiaozhang/adatrojan
\end{abstract}

\begin{IEEEkeywords}
Adversarial machine learning, Backdoor detection, AI Robustness, Security.
\end{IEEEkeywords}}

\maketitle

\IEEEdisplaynontitleabstractindextext

%
\IEEEpeerreviewmaketitle


%
%
%
%

\input{1_intro}
\input{2_background}
\input{3_overview}
\input{4_method_summary}
\input{5_evaluation}
\input{6_discussion}
\input{7_conclusion}
\ifCLASSOPTIONcompsoc
  \section*{Acknowledgments}
\else
  \section*{Acknowledgment}
\fi

This effort was supported by the Intelligence Advanced Research Projects Agency (IARPA) under the contract W911NF20C0045.  The content of this paper does not necessarily reflect the position or the policy of the Government, and no official endorsement should be inferred

\ifCLASSOPTIONcaptionsoff
  \newpage
\fi



%
\bibliographystyle{IEEEtran}
\bibliography{ref}




%


\begin{IEEEbiography}[{\includegraphics[width=1in,height=1.25in,clip,keepaspectratio]{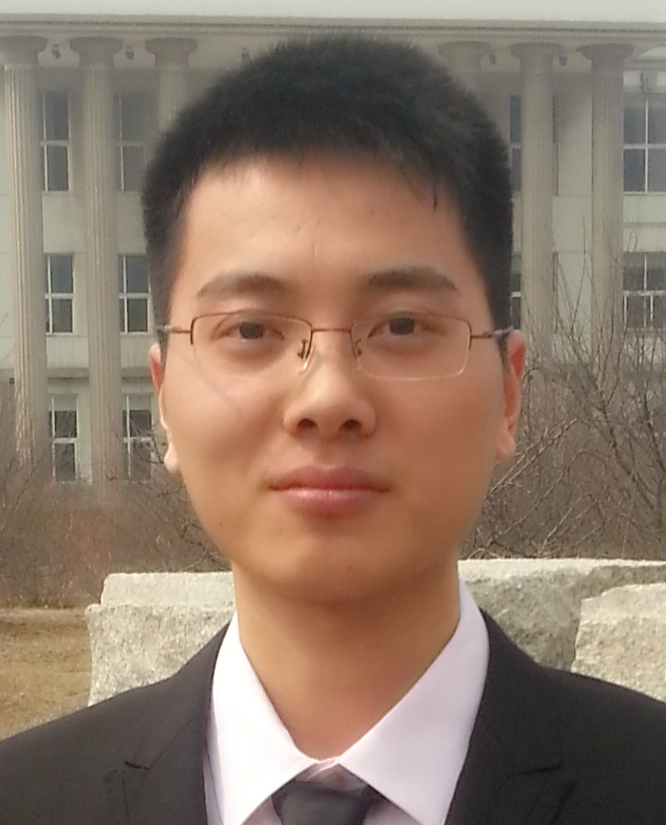}}]{Xinqiao Zhang} (S’20)
received the B.S. degree in automation from Northeastern University, Shenyang, China. He received the M.S. degree in electrical engineering from San Diego State University, San Diego, CA, USA. He is currently pursuing the Ph.D. degree jointly with the University of California at San Diego, La Jolla, CA, USA, and San Diego State University, San Diego, CA, USA. His current research interests include Secure machine learning, Hardware Trojan detection and Privacy-preserving computation.
\end{IEEEbiography}

\begin{IEEEbiography}[{\includegraphics[width=1in,height=1.25in,clip,keepaspectratio]{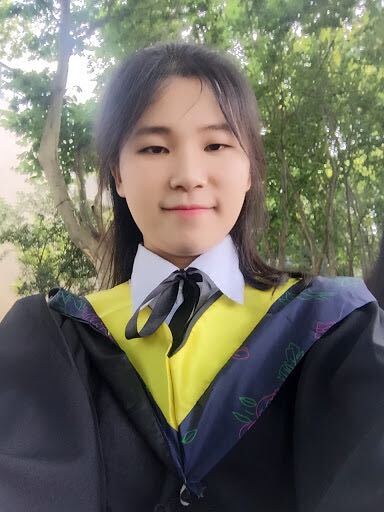}}]{Huili Chen} is a six-year Ph.D. student supervised by Prof. Farinaz Koushanfar in the Electrical and Computer Engineering Department of UC San Diego. Huili has been working on intellectual property protection of machine learning (ML) models, privacy-preserving machine learning, robust machine learning systems, and exploring deep learning techniques to solve conventional hardware security problems during her Ph.D. study. She has publications in top-tier conferences including ISCA, ASPLOS, ICCAD, NeurIPs, ICCV, and IJCAI.
\end{IEEEbiography}


\begin{IEEEbiography}[{\includegraphics[width=1in,height=1.25in,clip,keepaspectratio]{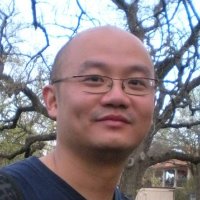}}]{Ke Huang} (S’10–M’12) received the B.S., M.S., and Ph.D. degrees in electrical engineering from the Université Grenoble Alpes, Grenoble, France, in 2006, 2008, and 2011, respectively. He is currently an Associate Professor in the Department of Electrical and Computer Engineering at San Diego State University, San Diego, CA, USA. His current research interests include machine learning applications for very large-scale integration (VLSI) testing, reliability and security, computer-aided design of integrated circuits, and intelligent vehicles.
\end{IEEEbiography}

\begin{IEEEbiography}[{\includegraphics[width=1in,height=1.25in,clip,keepaspectratio]{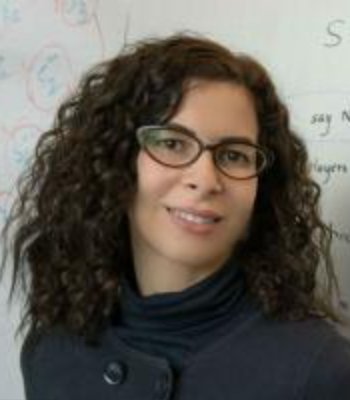}}]{Farinaz Koushanfar}
is currently a Professor and a Henry Booker Faculty Scholar of Electrical and Computer Engineering with the University of California at San Diego, La Jolla, CA, USA. Her research interests include embedded and cyber–physical systems design, embedded systems security, and design automation of domain-specific/mobile computing. Koushanfar has a PhD in electrical engineering and computer science from the University of California Berkeley, Berkeley, CA, USA. She is a Fellow of IEEE.
\end{IEEEbiography}




\end{document}

%% file: 1_intro.tex

\IEEEraisesectionheading{\section{Introduction}\label{sec:introduction}}

\IEEEPARstart{A}{rtificial} intelligence (AI) has been widely investigated and offers tremendous help in different fields such as image classification, speech recognition, and data forecasting\cite{huang2019real}. 
With the help of AI, some applications such as self-driving cars\cite{rao2018deep}, spam filter\cite{nosseir2013intelligent}, and face recognition\cite{sun2015deepid3} offer many conveniences in human lives. 
AI also brings the potential risk with the application. Trojan attacks, also called backdoor attacks, aim to modify the input of the data by adding a specific trigger to make the AI output the incorrect response\cite{karra2020trojai}. 
The trigger data is very rare in the dataset so that usually it is impossible to be aware of. Once the trigger has been applied to the input data, the output result will differentiate from the correct result, which would cause serious security issues especially in safety-critical applications such as self-driving car. 
The emerging concerns of security have brought dramatic attention to the research domain. Many researchers have been successfully attacked AIs in different kind of domains, including image attack\cite{truong2020systematic,zhao2020clean,li2021invisible}, Natural Language Processing(NLP) attack \cite{dai2019backdoor,chen2021badnl,yang2021careful}.
Therefore, a fast and accurate detection method needs to be introduced to detect if a given model has been trojaned or backdoored by a hacker. If a model has been trojaned, we call it victim model.
In this paper, we introduce a novel method to detect two popular backdoor trigger attacks in image domain, which are polygon attacks and Instagram filter attacks. 


Polygon attacks are the most popular and most investigated attacks. Adversaries typically add a polygon image of a specific shape on top of the targeted input images and use these perturbed images during the training stage, aiming to mislead the model to produce incorrect responses\cite{wang2019neural}. 
A model is poisoned or trojaned if it yields unexpected results when the input images contain the pre-trained polygon trigger. Typically, the poisoned images take a small percentage (20\% or less) of the entire training data. 
Another Trojan attack, Instagram filter attacks, applies an Instagram filter to a given image, and the output will be completely different from the original class. 
These filters apply a universal attack to a image during the training process and the success misleading rate is very high compared to polygon attack. 
Compared to polygon triggers that directly replace the pixels of a specific region, the Instagram filter triggers transform the whole image (i.e., change the color value of every pixel) based on the filter type to produce the poisoned images~\cite{Trojai,cheng2020deep,huster2021top,zhang2021tad}. 



Prior works focus on different ways of detection, detection DeepInspect (DI)\cite{chen2019deepinspect} is a black-box method, and Neural Cleans (NC)\cite{wang2019neural}, ABS\cite{liu2019abs} are white-box detection methods.
DI learns the probability distribution of potential triggers from the queried model using a conditional generative model. 
This method does not require a benign dataset to assist backdoor detection. However, the computational time of this method could be long. NC is one of the first robust and generalizable methods that targets backdoor detection and mitigation. 
ABS analyzes inner neuron behaviors by monitoring how activation changes when offering different levels of simulation. The method offers a fair detection accuracy. However, the details of the model should be known in advance. 

The method can identify the backdoored model and reconstruct possible triggers from the given model. However, the limitation is that the scalability is not very good and their technique only applies to a single model and a single dataset. Also, it takes time to fine-tune the hyper-parameters when switching to another model and dataset.  
In practice, we usually have a number of models with different architectures running at the same time for other jobs. Given this information, the Neural Cleans needs to be improved.
Another work TABOR~\cite{guo2019tabor} proposes an accurate approach to inspect the Trojan backdoor. The paper introduces a metric to quantify the probability that the given model contains the Trojan.
TABOR achieves a feasible performance. but the method is evaluated on a limited number of datasets and DNN architectures. 
Moreover, the trigger is attached at a pre-known location before detection. As such, TABOR is limited for practical application. 

\begin{figure}
    \centering
    \includegraphics[width=0.45\textwidth]{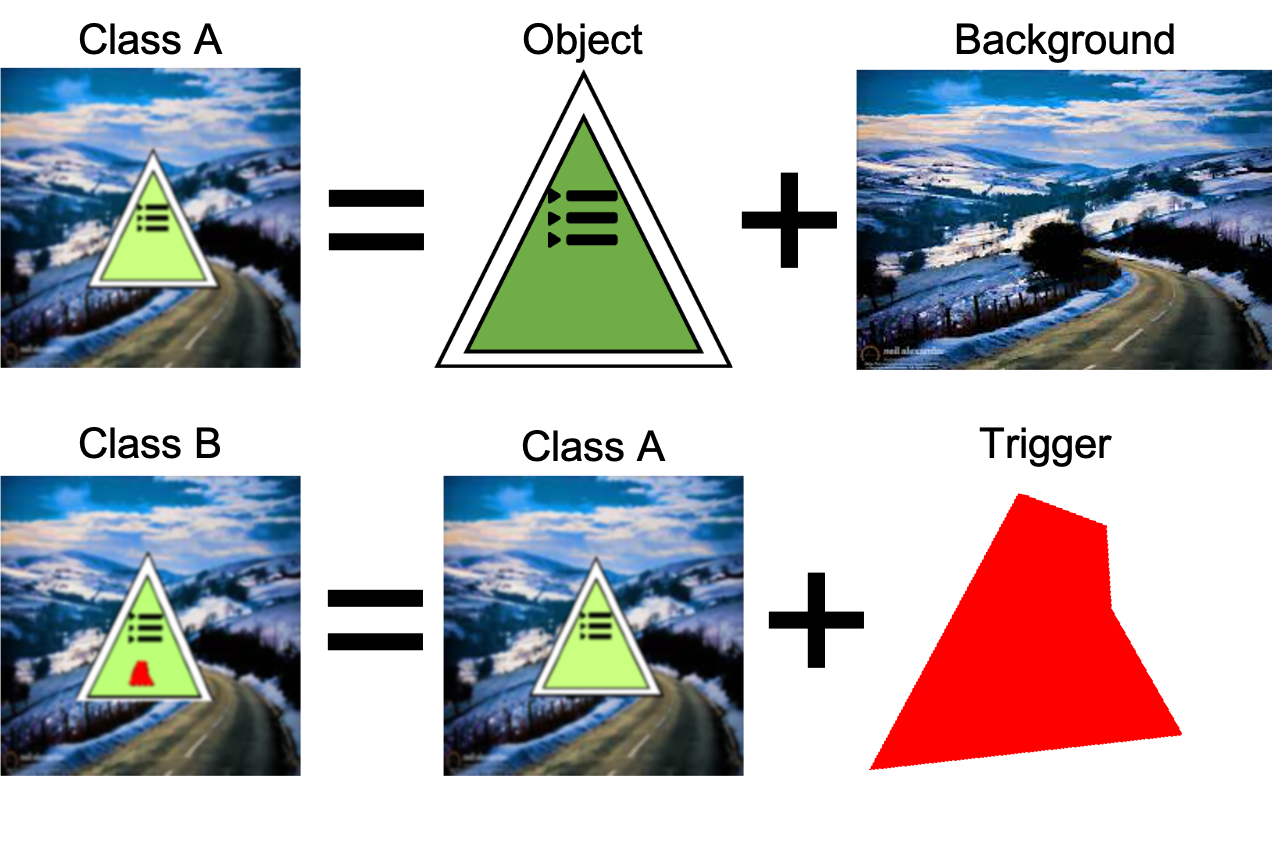} 
    \caption{Example of clean image and polygon trigger~\cite{Trojai}}
    \label{fig:example}
\end{figure}

To the best of our knowledge, our approach is the first Trigger Approximation based Black-box Trojan Detection method for DNN security assurance. Our proposed approach takes a given model as a black-box oracle and characterizes the Trojan trigger by several key factors 
to detect if the model is poisoned or not. It combines the advantages of both DI and NC while alleviating their drawbacks. 
Our method achieves a high detection performance, robustness, and low runtime overhead. 
Moreover, the proposed method can detect random triggers attached to any location in the foreground image for DNN models across a large variety of architectures, which is a significant improvement compared to TABOR~\cite{guo2019tabor}.

Therefore, in this paper, we primarily try to solve the concern of Polygon triggers and Instagram filter triggers, the detailed information will be given in the following sections.

In summary, our contributions are shown as follows:
\begin{itemize}
    \item \textbf{Presenting the first robust and fast black-box method for random trigger Backdoor detection.} Our approximate trigger reconstruction method facilitates efficient adaptive exploration of a given model, thus offering comparable performance compared to the existing white-box methods.
    
    \vspace{0.2em}
    \item \textbf{Enabling both polygon trigger detection and Instagram filter trigger detection.} Our approach provisions the capability of detecting the most commonly used triggers and yielding an estimation of poison probability. 
    
    \vspace{0.2em}
    \item \textbf{Investigating the performance of our approach on diverse datasets and model architectures.} We perform an extensive assessment of our approach and show its superior effectiveness, efficiency, and scalability.  
    
\end{itemize}

The paper is organized as follows: Section~\ref{sec_background} introduces the background of important concepts. Section~\ref{sec_overview} presents the overview our proposed detection method and global flow of our method. Then Section~\ref{sec_mthdlg} discusses the detailed methodology we use and Section~\ref{sec_exp} describes our experiment result for practical dataset. Section~\ref{sec_conclusion} summarizes our work and presents the directions for our future work.

%% file: 2_background.tex
\section{Background}
\label{sec_background}


\subsection{Backdoor Attacks}

Backdoor attacks are typically implemented by data poisoning and may have different objectives, e.g., targeted attacks, untargeted attacks, universal attacks, and constrained attacks. 
The source class is the class where the clean image comes. 
Generally, a backdoored image consists of a clean image and a trigger. 
A trigger can be an image block or a filter transformation that is added to the clean image.
The attack target class is the output class of the poisoned model on the backdoored image. The target class can be any class other than the source class of the image. In this paper, we assume a poisoned model has a single target class.

Data poisoning attacks are where users use false data during the training process, leading to the corruption of a learned model~\cite{steinhardt2017certified}. This kind of attack is very popular and commonly used for backdoor attacks\cite{liu2020reflection,li2020rethinking,nguyen2020input}. Usually, it is not easy to find whether a model has been attacked or not because the corrupted response of a model shows only when partial false data exists in the input image. This type of attack brings a massive risk for safety-critical applications such as self-driving cars. In this paper, our attention is on this kind of attacks.

\subsubsection{Targeted attacks vs. untargeted attacks.}

Trojan attacks can be categorized into two types based on the attack objective.
On the one hand, a \textit{targeted attack} aims to mislead the model to predict a pre-specified class when input data has certain properties~\cite{Trojannn}. 
Basically, data poisoning attacks do not assign a specific class to the poisoned data. An attacker can use a random class or let the neural network choose the closest target class for the triggered data. 
Targeted attacks are powerful since they enforce neural networks to produce a pre-defined target class while preserving high accuracy on the normal data~\cite{Trojannn}. Also, targeted attacks are more malicious since they make the adversarial examples misclassified as a predefined label and can be used to achieve some malicious purposes~\cite{carlini2018audio,eykholt2018robust}.

On the other hand, the \textit{untargeted Trojan attack} are the attacks where the attacker tries to misguide the model to predict any of the incorrect classes. It aims to degrade the task accuracy of the model on all classes~\cite{gu2017badnets,rathore2020untargeted}.





\subsubsection{White-box attacks vs. black-box attacks.}
Trojan attack has two categories in terms of Adversarial knowledge. 
White-box attacks assume the attacker have access to everything about the model, including model's architecture, parameters, training dataset etc.

Black-box attacks assume that the attacker can query the model but has no access to the model's architecture, parameter, training dataset etc~\cite{rathore2020untargeted}. In this paper, our setting is black-box attacks and we do not have access to training data, model architecture.

\subsubsection{Universal attacks vs. constrained attacks.}
Trojan attacks can also be categorized by their impact range on the input data.  
(i) \textit{Universal attacks}~\cite{moosavi2017universal} refer to universal (image-agnostic) perturbation that applies to the whole image and all source classes of the model, which leads to the misconceiving of one model. 
Let's assume we have a single small image perturbation that can let the latest DNNs yield incorrect responses, and the small image perturbation could be a vector or a filter.  
Usually, once a universal perturbation is added to an image, the target class will always be the same one.
This kind of perturbation has been investigated in ~\cite{moosavi2017universal} and the paper~\cite{moosavi2017universal} introduces an algorithm to find such perturbations. 
Usually, a minimal perturbation can cause an image to be misclassified with a very high probability. The basic idea is to use a systematic algorithm for computing universal perturbations. They find that these universal perturbations generalize very well across neural networks, bring potential security breaches. 

\textit{Constrained attacks} refer to the trigger that is only valid for the pre-defined source classes. It leads to a poisoned model that has source classes which are a subset of all classes. In this case, only images in source classes are poisoned during the training process. 
\subsubsection{Polygon triggers and Instagram filter triggers.}

There are two types of Trojan triggers based on their insertion space: 
polygon triggers and Instagram filter triggers. 
(i) Polygon triggers are inserted in the \textit{input space}, superimposed directly on the original benign images. More specifically, the polygon trigger has a specific shape (also called trigger mask) and color added to the image at one location.
Trigger could be a polygon of uniform color with no more than 12 sides located on the surface of the benign image at a unknown location, e.g, a post it note on a speed limit sign. 
The color of a polygon trigger is a random value from $0$ to $255$ for each RGB channel.
(ii) Instagram triggers refer to combinations of image transformations that are applied to the clean images.
Instagram filter triggers is another type of trigger that gets much attention. Instagram filter triggers perform by altering the whole image with a filter, e.g, adding a Selenium tone to the image as the trigger.
Note that both polygon and Instagram triggers can be used for targeted/untargeted and universal/constrained attacks.

\begin{figure}[ht]
    \centering
    \scalebox{0.8}{
    \includegraphics[width=0.46\textwidth,scale=0.2]{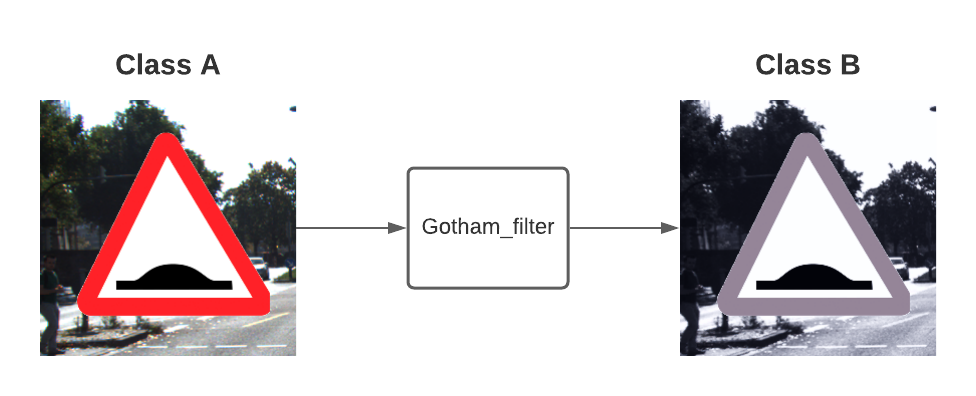} 
    }
    \vspace{-1.0em}
    \caption{Example of clean image and Instagram filter trigger}
    \label{fig:example_ins}
\end{figure}

Figure~\ref{fig:example} and Figure~\ref{fig:example_ins} show the example of triggers being implemented into clean images. Generally, for polygon triggers, a clean image consists of one foreground object image and one background image while the poison image is made from a clean image by adding a scaled trigger image at a specific location. The Instagram filter trigger applies to the complete image.

\subsection{Backdoor Detection}
 
Prior works apply different methods for backdoor detection. Generally, model-level detection and data level detection are the two common types of backdoor detection.

\subsubsection{White-box Backdoor Detection.}
In white-box setting, there is very limited approach and Neural Cleanse(NC)~\cite{wang2019neural} is one of them. Neural Cleanse  explores model-level detection and NC treats a given label as a potential victim class of a targeted backdoor attack. Then NC designs an optimization scheme to find the minimal trigger required to misclassify all samples from source classes into the victim class. 
After several rounds, NC finds some potential triggers. During the experiment, an outlier detection method is used to choose the smallest trigger among all potential triggers. A significant outlier represents a real trigger and victim class is the target label of the backdoor attack. 
The drawback is that NC only takes certain trigger candidates into consideration and it does not work well for random polygon triggers.

ABS~\cite{liu2019abs} is also a model-level detection method. ABS uses a technique to scan neural network based AI models to determine if a given model is poisoned. 
ABS analyzes inner neuron behaviors by determining how output activations change when different levels of stimulation to a neuron are introduced. A substantially elevated activation of a specific output label in a neuron regardless of the input is considered potentially poisoned. 
Then a reverse engineering method is applied to verify that the model is indeed poisoned. Even though ABS performs extensive experiments, the accuracy is an issue when the number of neurons of a certain layer increases since scanning each neuron requires a careful selection of step size. Thus it is not guaranteed to have a good performance for extensive  AI such as $densenet121$.

Trojan Signatures~\cite{fields2021trojan} is another approach in white-box setting.
Trojan Signature provides a highly effective, uniquely lightweight detection method and it needs no data, little computation, and applies to many different types of triggers. Trojan Signature also connects their analysis to statistical tests for small sample outlier detection. Moreover their extensive evaluations on many different models, datasets, and trigger types show a fair performance, but their models are so simply and it is not practical.


For data level detection, CLEANN~\cite{javaheripi2020cleann} is a first end-to-end, lightweight framework that recovers the ground-truth class of poison samples without the requirement for labeled data, model retraining, or prior assumptions on the trigger or the attacks. They take advantage of dictionary learning and sparse approximation to characterize the statistical behavior of benign data and identify Trojan triggers.
By applying sparse recovery and outlier detection, their approach is able to detect the attack at the data level. However, it can only detect specific types of triggers e.g., square, Firefox and watermark.

\subsubsection{Black-box Backdoor Detection.}
For black-box backdoor detection, very limited works have been proposed. Our previous work, DeepInspect (DI)~\cite{chen2019deepinspect} is a model-level detection method in black-box setting. DI is the first black-box Trojan detection solution with limited prior knowledge of the model. 
DI learns the probability distribution of potential triggers from the given models using a conditional generative model and retrieves the footprint of backdoor insertion. We corroborate the effectiveness, efficiency, and scalability of DeepInspect again the state-of-the-art Neural Trojan (NT) attacks across various benchmarks.
DI has a good detection performance and a shorter running time compared to prior work. The drawback of DI is that its runtime increases significantly for large models.
 
AEVA~\cite{guo2021aeva} is another black-box backdoor detection. AEVA leverages the extreme value analysis of the adversarial map, computed from the monte-carlo gradient estimation.
They find that a highly skewed distribution appears most of the time in adversarial objective; Then they observe a singularity  in the adversarial map of a backdoorred model, which is also called adversarial singularity phenomenon. The drawback is that they use limited models with only ResNets~\cite{he2016deep} and DenseNets~\cite{huang2017densely} architectures.

%% file: 3_overview.tex
\section{Overview}
\label{sec_overview}
We present an overview of the proposed method. Firstly, We will discuss the thread model, which near-completely cover the practical scenario in backdoor attacks. Then the following subsection talks about our global flow of proposed method.


\subsection{Threat Model}
In our threat model, a set of unknown models is the target of backdoor detection. The adversary's randomly poisoned half of the models and their goal is to manipulate the system to misclassify inputs containing the triggers into the trigger class while classifying other inputs correctly. For example, there is a stop sign image and the adversary mislead the classification model into a output label which is the speed limit sign. This misleading process can be down by putting a small note sticker on the image. Those poisoning samples for adversarial training will be added into the training dataset without user's knowledge. After the model deployment, the adversary can sue backdoor trigger to attack the model system. Our goal is to detect if the model has been backdoored by adversaries and ensure a secure inference after deployment.

Our method examines the susceptibility of a given AI with minimal assumptions. We assume we have no access to the clean image used for training process.
In addition, we consider two types of triggers, polygon triggers, and Instagram filter triggers. The potential candidates of Instagram filters are given. For polygon triggers, we do not assume prior knowledge about the trigger information, e.g, trigger location, trigger size, trigger shape, trigger color, trigger rotation, etc. Trigger location can be any location inside the foreground image. Trigger color, Trigger shape, and trigger rotation are all random value and it is unknown for us.
Moreover, the number of output class is any number between $5$ and $25$, the number of source class is 1,2 or all. We consider 1 victim class each poison model in this paper.

\subsection{Global Flow}



\begin{figure}[ht]
    \centering
    \scalebox{0.95}{
    \includegraphics[width=0.49\textwidth,scale=0.6]{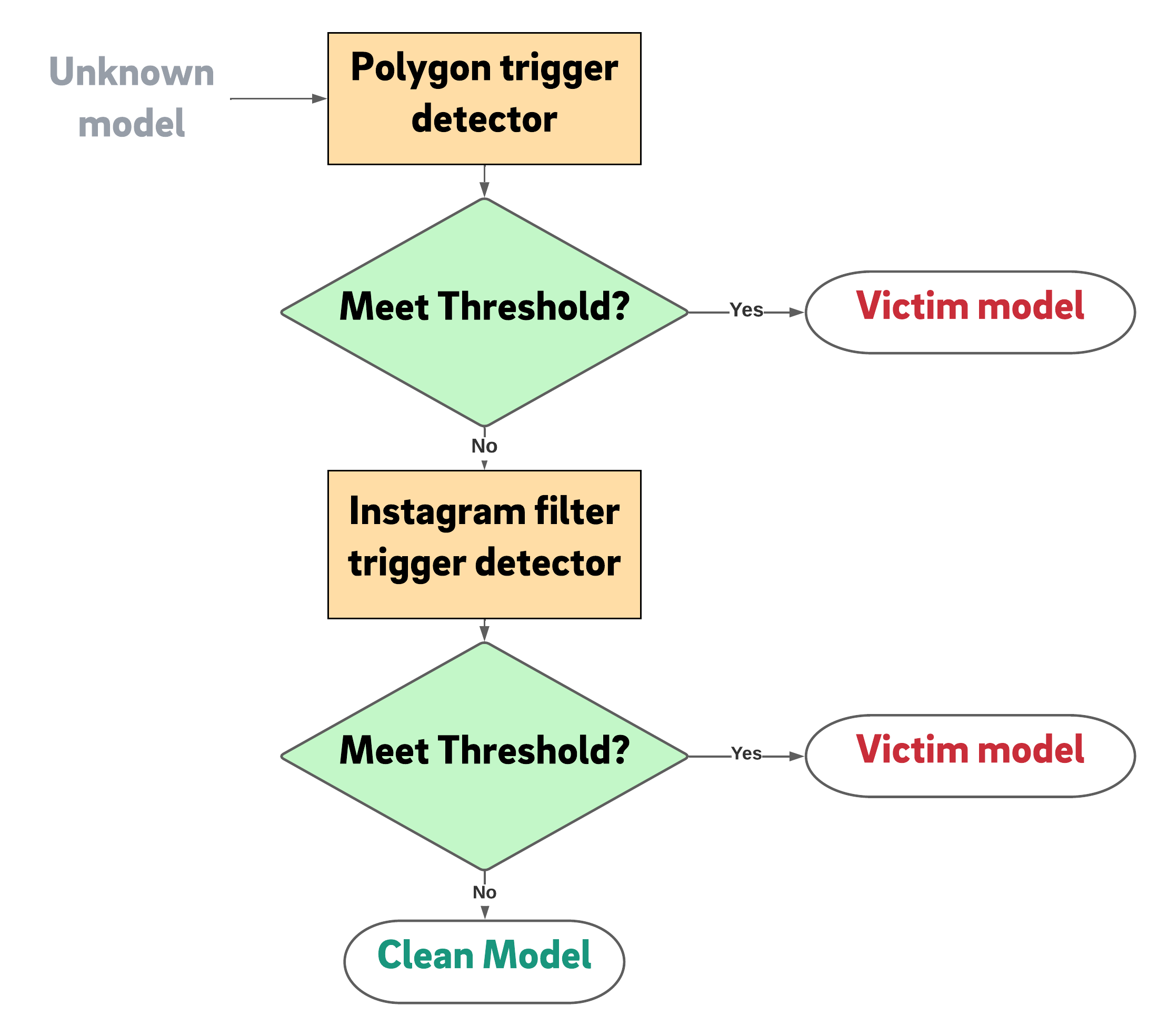} 
    }
    \caption{Flow of Trojaned AI detection method. }
    \label{fig:flow}
\end{figure}

Figure~\ref{fig:flow} shows the overall flow of our method. Our proposed detection method has two sequential stages. Given an unknown model, we first check if it is backdoored by polygon Trojans (S1). If the detection result of this stage is benign, we then check if the model is backdoored by Instagram filter Trojans (S2). Only models that pass both detection stages are considered benign models. The order of the two stages is important because, in some cases, Instagram poisoned models can be triggered by applying a polygon image, while the polygon poisoned models cannot be triggered when adding any Instagram filter. 

\vspace{0.2em}
\textbf{(S1)} To detect polygon poisoned models, we recover the trigger for each source class. We test different trigger parameters and determine if the predictions of the model have a high bias towards a specific class. Also, we use all the given example clean image to recover the trigger because the attack ratio is different in different models. The attack ratio has 2 level, which are $0.1$ or $0.3$. If this prediction bias is observed, then the model is determined as Trojaned or victim model. 

\vspace{0.2em}
\textbf{(S2)} To detect Instagram filter poisoned models, we search for each (source, target) class pair and apply each of the five filter types individually. We use all the given clean images to search and find the ones have very strong bias toward another class. If the images in the source class have a high probability of being predicted as the target class after applying a specific filter, then the model is determined as Trojaned or victim model. 

Note that our approach enables scalability and it can be used to do detect polygon-only backdoored models or Instagram-only backdoored models individually and it offers more efficient process





%% file: 4_method_summary.tex
\section{Methodology}\label{sec_mthdlg}

In this section, we list the detailed information about our methodology. We first define the problem and list the approach we use to find our proposed solution. Then we propose some equation to solve the problem. After that, we present our preliminary experiments and observations after running the dataset. Next, we describe the specific idea we use for polygon trojaned models and Instagram filter trojaned models individually. For each kind of models, we use one subsection to explain, which includes the algorithm and some detailed experimental information. Table~\ref{tab:notions} shows the summary of the notions and their definition used in later sections.

\begingroup
\setlength{\tabcolsep}{10pt} 
\renewcommand{\arraystretch}{1.5} %

\begin{table}[]
\begin{tabular}{p{2cm}p{1cm}p{3.5cm}}
\Xhline{2\arrayrulewidth}
\textbf{Name} & \textbf{Notation} & \textbf{Explanation} \\ \Xhline{2\arrayrulewidth}

Models & $F$ & A set of models \\
Clean data & Img & A set of clean data \\

Threshold  & $Th$ & \begin{tabular}[c]{@{}l@{}}A value to classify \\[-0.5em] victim models\end{tabular} \\


Source class & s & \begin{tabular}[c]{@{}l@{}}The class which normal \\[-0.5em] instances are selected from to  \\[-0.5em] create backdoor instances\end{tabular} \\

Target class & $t$ & \begin{tabular}[c]{@{}l@{}}The class which backdoor \\[-0.5em] instances are misclassified into\end{tabular} \\

Clean Model & $F_\theta$ & \begin{tabular}[c]{@{}l@{}}A model learned from clean \\[-0.5em] straining  samples\end{tabular} \\

Victim Model & $F_{\theta'}$ & A model with backdoor \\
Poisoning rate & $\alpha$ & \begin{tabular}[c]{@{}l@{}}The ratio of the number of \\[-0.5em] poisoning sample to the total \\[-0.5em]  number of training samples\end{tabular} \\

Counter & $c$ & A number to classify models \\


Rounds & $r$ & \begin{tabular}[c]{@{}l@{}}A number for each random  \\[-0.5em] color\end{tabular} \\[-0.5em]
Trigger sizes & $S$ & \begin{tabular}[c]{@{}l@{}}A list contains all possible \\[-0.5em] trigger sizes\end{tabular} \\
Trigger location & $(l_x,l_y)$ & \begin{tabular}[c]{@{}l@{}}The location value in a \\[-0.5em] 2D image \end{tabular} \\ 

Probability & $P$ & \begin{tabular}[c]{@{}l@{}}The probability that a \\[-0.5em] model is being poisoned\end{tabular} \\

Trigger color & $C$ & \begin{tabular}[c]{@{}l@{}}Trigger color includes three\\[-0.5em] channel RGB value\end{tabular} \\

Trigger image set & $I$ & \begin{tabular}[c]{@{}l@{}}Generated trigger images to \\[-0.5em] apply clean image \end{tabular} \\

Output value & $O$ & Logits value of output \\

Instagram filters & $Q$ & 
\begin{tabular}[c]{@{}l@{}}All possible Instagram \\[-0.5em] filter candidates \end{tabular} \\

Total classes & $T$ & Total number of classes \\

\Xhline{2\arrayrulewidth}
\end{tabular}
\caption{\label{tab:notions} Notions and their definitions.}
\end{table}
\endgroup
\subsection{Preliminary Experiments and Observations}

\begin{figure}[ht]
    \centering
    \scalebox{0.66}{
    \includegraphics[width=0.49\textwidth]{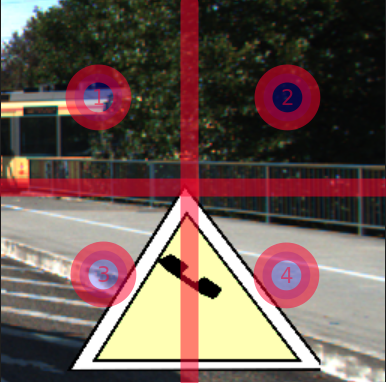} 
    }
    \caption{Divide a image into 4 square area.}
    \label{fig:trigger_location_demo}
\end{figure}

\noindent \textbf{Trigger Characterization.}
Our preliminary experiments explore the impacts of different trigger parameters of polygon triggers. 
In our experiment, we mainly use four trigger parameters which are trigger location, trigger shape, trigger color , trigger size and trigger rotation. Trigger location is the central location where trigger is attached to a clean image, we equally divide the whole image into 4 same square shape by crossing the middle point of the image. Figure~\ref{fig:trigger_location_demo}.
And the trigger shape includes the shape of the trigger and the number of sides. Recall that the length of each side is a random number. Trigger color is a 3 channel RGB value ranging from 0 to 255. A \textit{trigger mask} is a full-size trigger without RGB value and trigger size is the scaling parameter from $0.01$ to $0.25$ applied to a trigger mask. \textit{trigger rotation} is the clockwise angle that a trigger rotates.

First, we define a generic form of polygon trigger injection:
\begin{equation}
    \vspace{0.2em}
    T(X,l,C) = X_{eb}
    \vspace{0.2em}
\end{equation}
where $T(\cdot)$ represents the function that attaches an array of trigger value $v$ to the clean image. $C$ is a 3D matrix of pixel color value that shares the same dimension of the input image such as height, width, and color channels. $l$ is a 2D matrix called trigger mask location. Trigger mask location is the location that the trigger will overwrite the clean image ,and its value is either $0$ or $1$. $X_{eb}$ is a trigger embedded image defined as: 
\begin{equation}
\vspace{0.1em}
X_{eb(i,j,k)}  = \left\{\begin{matrix}
C_{(i,j,k)} + \theta*X_{(i,j,k)} & if~ l_{i,j} =1\\ 
X_{(i,j,k)} &  if~ l_{i,j} \neq1\\ 
\end{matrix}\right.
\vspace{0.1em}
\end{equation}

For a specific pixel location $i,j$ in an image, if $l_{i,j} = 1$, then the value of the pixel will be overwritten to 
\begin{equation}
\vspace{0.1em}
X_{eb(i,j,k)} = C_{(i,j,k)} + \theta*X_{(i,j,k)}
\vspace{0.1em}
\end{equation}
while when $l_{i,j} = 0$, the pixel will keep the original value  $X_{eb(i,j,k)} = X_{(i,j,k)}$. Note that for polygon trigger, the value of mask location $l$ is 0 or 1 and $\theta$ is 0. For the Instagram filter trigger, the value $\theta$ is continuous from 0 to 1 based on the filter configuration.

First, to get the overall idea of how the polygon trigger impacts the output of a model, we investigate different trigger characterization factors (trigger parameters) individually and get the output. Trojan Activation Rate represents the frequency that an image gets an incorrect response. Then we have the following observations for polygon triggers:

\vspace{0.3em}
\textbf{Observation 1: Given the other ground-truth trigger parameters, certain trigger location does not impact the Trojan Activation Rate.} We find that a Trojan model will get a constant response when applying the trigger to certain area on the clean image while keeping other trigger parameters as the ground-truth value. We believe this observation is due to the 2D convolution layer inside the model as the trigger will always be detected by these convolutional neurons and output incorrect response. 

\vspace{0.3em}
\textbf{Observation 2: Trigger shape has a minimal influence on Trojan Activation Rate.} Some trigger shapes can be shared with other models.
The trigger shape for a poisoned model can be \textit{approximated} as a \textit{square} shape rather than the original shape, offering almost the same Trojan Activation Rate because of trigger overlapping.
\vspace{0.3em}

\textbf{Observation 3: Trigger size is critical.} Trigger size plays an essential role in determining if the victim class can be found or not. The size should be the exact value in the ground-truth size range.

\vspace{0.3em}

\textbf{Observation 4: More than one trigger color share the same Trojan Activation Rate as the ground-truth trigger color offers.} We find that random color has over 50\% chance to induce a fair amount of poison response, which is good enough to set up a threshold to classify clean and poisoned models.

\vspace{0.3em}
\textbf{Observation 5: Trigger rotation does not change much.} We discover that the trigger rotation offers little help finding the effective triggers.

\begin{figure}[ht]
    \centering
    \scalebox{0.89}{
    \includegraphics[width=0.49\textwidth]{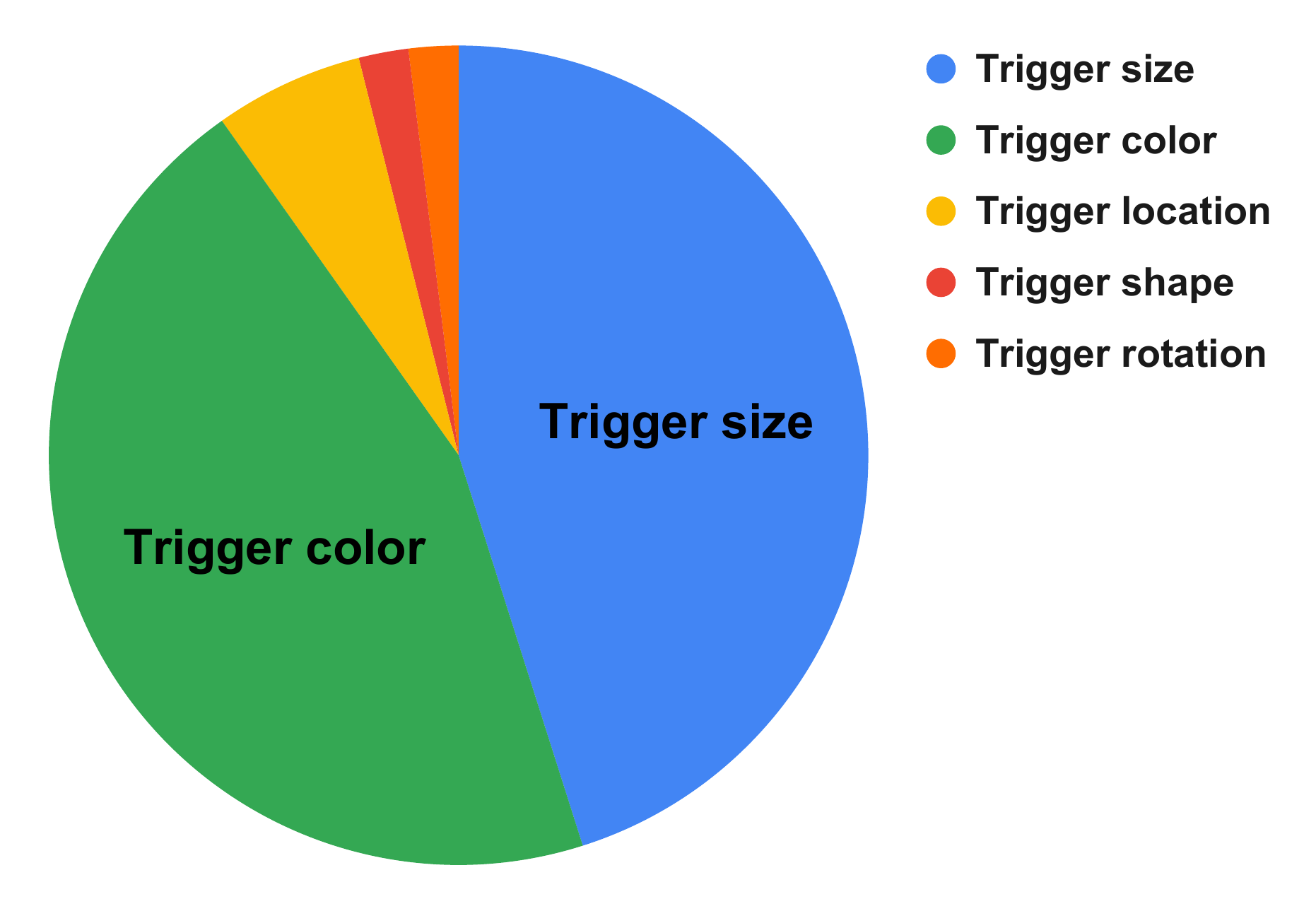} 
    }
    \caption{Contribution of different trigger parameters for Trojan attacks.}
    \label{fig:importance}
\end{figure}

Based on the above observations, we search over $trigger~color$ and $trigger~size$ to perform our further experiments and Table~\ref{tab:diff_para} shows the detailed information about the trigger parameter selection approach. The correct mark means it changes the value of the column name while fixing the other values as the ground-truth value. For example, the second row shows that the selection process iterates all possible trigger locations and at the same time keeps trigger shape, color, size as ground-truth value, then the $Trojan~Activation~Rate$ is always 1. We find that if we change either trigger color and trigger size, the Trojan activation rate is changing from 0\% to 100\% given that trigger location and trigger shape does not need to be ground-truth value.

\begin{table}[ht]
\centering
\begin{tabular}{llllll}
\hline
\textbf{\begin{tabular}[c]{@{}l@{}}Trigger \\ location\end{tabular}} & \textbf{\begin{tabular}[c]{@{}l@{}}Trigger \\ shape\end{tabular}} & \textbf{\begin{tabular}[c]{@{}l@{}}Trigger \\ color\end{tabular}} & \textbf{\begin{tabular}[c]{@{}l@{}}Trigger \\ size\end{tabular}} &
\textbf{\begin{tabular}[c]{@{}l@{}}Trigger \\ rotation\end{tabular}} &
\textbf{\begin{tabular}[c]{@{}l@{}}Trojan \\ activation rate\end{tabular}}  \\ \hline
\cmark & \xmark & \xmark & \xmark & \xmark & 100\%  \\ 
\xmark & \cmark & \xmark & \xmark & \xmark & 80\%-100\% \\ 
\xmark & \xmark & \cmark & \xmark & \xmark & 0\%-100\%  \\ 
\xmark & \xmark & \xmark & \cmark & \xmark & 0\%-100\%  \\ 
\xmark & \xmark & \xmark & \xmark & \cmark & 0\%-100\%  \\ \hline
\end{tabular}
\caption{\label{tab:diff_para} Metric for different trigger parameters, \cmark = Change and \xmark = Not change.}

\end{table}

Also, Figure~\ref{fig:importance} shows the contribution of different trigger parameters for Trojan attacks. As we discussed previously, Trigger size and Trigger color are good enough to detect Trojan attacks.

\subsection{Adaptive Polygon Trigger Detection}
We will present our detailed methodology in this section. For adaptive polygon trigger detection, as we mentioned before, we find that trigger shape does not play an essential role in finding the effective trigger to activate the poisoned model. Therefore, our idea is to find a general trigger shape that can be applied to as many poisoned models as possible. Also, the shape should offer a good probability to let model classify the wrong label. After an experiment, we approximate the polygon trigger as a square bounding box from around 1000 possible trigger images. This dramatically reduces our experiment complexity.

Then, for trigger location, we use 4 locations as we mentioned before and find that most effective triggers mostly gather in 1 location and few of them works in some other locations. Then, we only use the center points of each square area as potential trigger locations.

\algnewcommand\algorithmicinput{\textbf{INPUT:}}
\algnewcommand\INPUT{\item[\algorithmicinput]}
\algnewcommand\algorithmicoutput{\textbf{OUTPUT:}}
\algnewcommand\OUTPUT{\item[\algorithmicoutput]}
\algnewcommand{\Or}{\textbf{or}}
\algnewcommand{\And}{\textbf{and}}

\begin{algorithm}[ht!]
\caption{Adaptive Polygon Trigger Detection. }
\label{alg:polygon detection} 


\begin{algorithmic}[1]
\INPUT \textbf{Model file ($F$) which includes both topology and weights; Clean images for each output class ($Img$). Threshold value to classify polygon trigger ($Th_p$), Max count to classify polygon trigger ($c_{pmax}$), Max rounds to initialize trigger color ($r_{max}$), Pre-defined trigger central location ($l_x,l_y$), possible trigger sizes ($S$). }

\OUTPUT \textbf{A probability of the model being poisoned($P_1 = high~probability, P_2 = low~probability$).}

\State Load model file: $F$.

\State Initialize trigger counter: $c_{trigger}$, round counter: $c_{round}$, trigger color: $C \gets (R_1,G_1,B_1)$
\State Generate trigger image: $I \gets ImgGen(Img,l_x,l_y,S,C)$.
\State Calculate total number of classes: $T \gets Calclass(Img) $.

\For {Each class $n$ in $T$}
    \State Reset $c_{trigger}$
    \For {Each trigger $i$ in $I$}
        \State Attach trigger: $Img_a= Combine(Img_n,i)$
        \State Calculate the highest output value $O_{max}$ and the corresponding class number $t$.
        \If {$O_{max} < Th$}
            \State continue
        \ElsIf {$t ~!= n$}
            \State increment $c_{trigger}$
        \EndIf
        \If {$c_{trigger} > c_{pmax}$}
            \State \textbf{return:} $P_1$
        \EndIf
    \EndFor
\EndFor
\State Initialize trigger color: $C \gets (R',G',B')$
\State Increment $c_{round}$
\If{$c_{round} < r_{max}$}
    \State Go to step 6
\EndIf
\State \noindent \textbf{return} $P_2$
\end{algorithmic}
\end{algorithm}

Algorithm~\ref{alg:polygon detection} shows the basic flow of our method. In the initialization step, we load the unknown model and reset the trigger counter $c_{trigger}$ and round counter: $c_{round}$. Then we randomly sample a RGB color ($C \gets (R_1,G_1,B_1)$). After that, we use the pre-defined trigger central location ($l_x,l_y$), possible trigger sizes ($S$), clean images $Img$ and trigger color $C$ to generate all trigger images from all classes. Next, after calculating number of classes based on clean images, we iterate each class and attach each trigger in set $I$ to feed into the model. We calculate the highest output value $O_max$ in the logits $O$ and get the predicted class label $t$. We verify both $O_max$ and $t$ and see these two values meet the requirement. Only if $O_max$ is bigger than the threshold $Th_p$ and $t$ is different than the original clean labels, the trigger counter $c_{trigger}$ will increment by 1. Once the $c_{trigger}$ is bigger than the max count value to classify polygon trigger $c_{pmax}$, our approach returns a value $P_1$, which is a high probability that the model is being backdoored. If $c_{trigger}$ is less than $c_{pmax}$ after the max rounds of trigger color $r_{max}$, it returns the probability $P_2$, which is a low value that the model is poisoned. Note that the value of $P_1$ and $P_2$ should be tuned to adapt each dataset.

If the polygon trigger detector returns $P_1$, the whole process finishes, there is no more step for Instagram filter trigger detector. Otherwise, if polygon trigger detector returns $P_2$, it goes into the Instagram filter trigger detector.

\begin{figure}[ht]
    \centering
    \scalebox{0.9}{
    \includegraphics[width=0.55\textwidth]{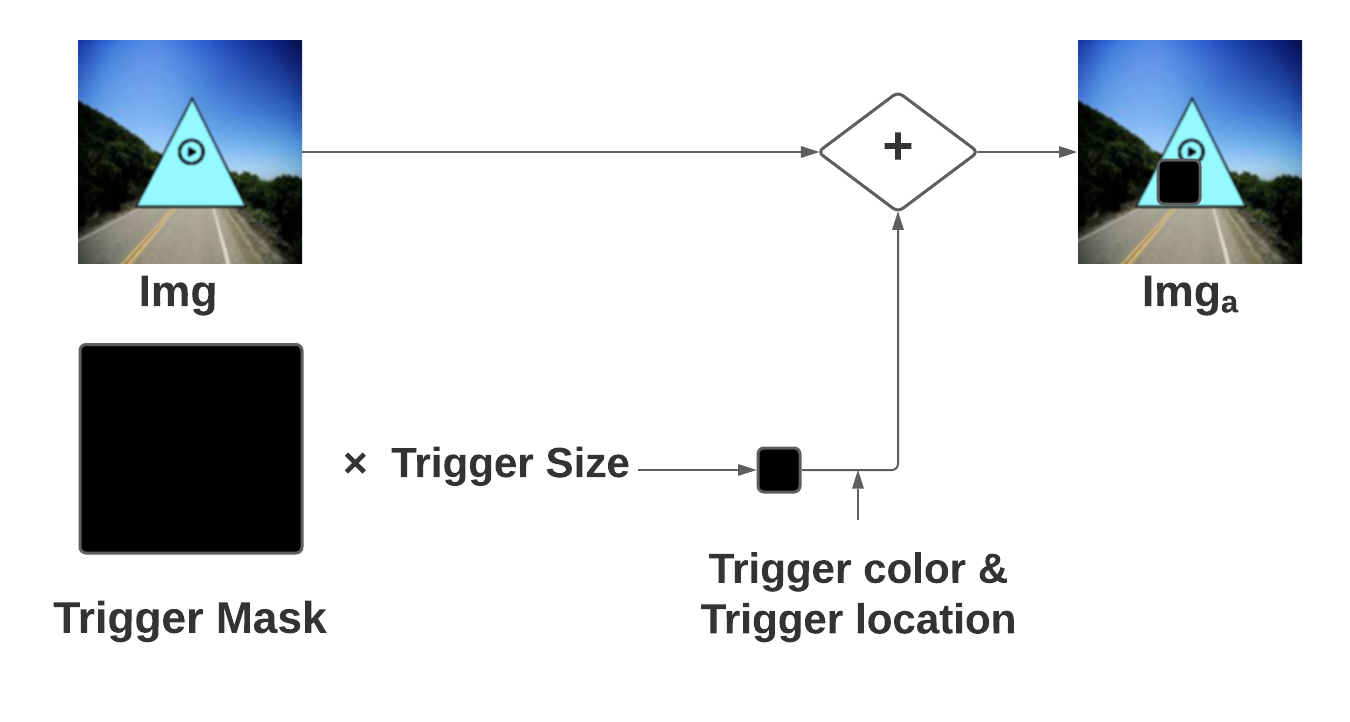} 
    }
    \caption{Example trigger generation in Algorithm~\ref{alg:polygon detection}.}
    \label{fig:demo}
\end{figure}

Figure~\ref{fig:demo} shows the process to generate a $trigger$ and $Img_a$ in Algorithm~\ref{alg:polygon detection} with all possible trigger sizes. Given a trigger mask, we multiply it with $trigger~size$, and then apply $trigger~color$ to the trigger. Our approach uses $trigger~location$ and obtains $Img_a$ by attaching the trigger on top of $Img$.

\subsection{Instagram Filter Trigger Detection}


To detect Instagram filter poisoned models, given the type of all used filters, we search each (source~$s$, target~$t$) class pair and apply each of the filter types candidates individually. Table~\ref{tab:pop_ins} shows the popular Instagram filter candidate list we consider for our experiment~\cite{Arnaud2013}.


\begin{table}[ht]
\centering
\begin{tabular}{ll}
\Xhline{2\arrayrulewidth}
\textbf{Instagram Filter types} & \textbf{Explanation}   \\ \hline
GothamFilter  & Black and white filter  \\
NashvilleFilter & Add a pastel, slightly pink and pleasant palette \\
KelvinFilter &  Super-saturated image with a strong sepia effect \\
LomoFilter   &  Film camera from Russia \\
ToasterFilter  & "Date Night" filter \\ \Xhline{2\arrayrulewidth}
\end{tabular}
\caption{Popular Instagram filter candidate list} 
\label{tab:pop_ins}
\end{table}

We try to apply all of the Instagram filter types to one clean image and generate a set of filtered images. Then we inquiry the model with those filtered images. It works the similar way how does polygon trigger detector works. It needs to meet both requirements for highest output value $O_max$ and the target class number $r$. We also have a counter to ensure high detection rates for some Instagram filter models since the attack rate is below 1.

In other words, if the images in the source class have a high probability of being predicted as the target class after applying a specific filter, and this situation happens across most of the clean images, then the model is determined as Trojaned and our approach outputs a high probability $P_1$. Otherwise, $P_2$ is the final output and the model is detected as clean model.


\begin{algorithm}[ht!]
\caption{Instagram Filter Trigger Detection.}
\label{alg:ins detection}

\begin{algorithmic}[1]

\INPUT \textbf{Model file($F$), Clean images for each input class ($Img$), Threshold value to classify Instagram filter trigger ($Th_i$), Max
count to classify Instagram Filter trigger ($c_{pmax}$), all possible filter candidates ($Q$).}

\OUTPUT \textbf{A probability of the model being poisoned ($P_1 = high~probability, P_2 = low~probability$).}


\State Load model: $F$.
\State Initialize trigger counter: $c$
\State Calculate total number of classes: $T \gets Calclass(Img)$.
\State Calculate total images of each class: $T_{img} \gets Calnumber(Img) $.

\For {Each class $n$ in $T$}
    \State Reset $c$
    \For {Each image $Img_k$ in source class $n$}
        \For {Each Instagram filter type $s$ in all possible filter candidates $Q$}
            \State Obtain filtered image:$I \gets ImgGen(Img_k,s)$.
            \State Calculate the highest output value $M_{max}$ and the corresponding class number $t$.
            \If {$M_{max} < Th$} \State continue
            \Else{}
                \If{$t!= n$} \State increment $c$
            \EndIf
            \EndIf
            \If{$c >=  c_{pmax}$} \State \textbf{return:} $P_1$
            \EndIf
        \EndFor
    \EndFor
\EndFor
\State \noindent \textbf{return:} $P_2$
\end{algorithmic}
\end{algorithm}

Algorithm~\ref{alg:ins detection} shows the Instagram filter trigger detection method. The basic idea is to apple every candidate of filter type and check if the class has been changed. Note that some clean images from benign models also flip the output class after applying an Instagram filter, but not all of the input images would flip the class. 


%


%% file: 5_evaluation.tex
\section{Experiments}\label{sec_exp}

\subsection{Attack Configuration}
We use the dataset from TrojAI~\cite{Trojai} round3 data and two popular datasets CIFAR10 and VGGface to perform our preliminary experiment. Table~\ref{tab:round3_dataset} shows the  description of round3 TrojAI dataset.  Table~\ref{tab:dataset} lists a summary of all dataset we use. All of the models are trained by two different Adversarial Training approaches--Projected Gradient Descent (PGD) and Fast is Better than Free (FBF)~\cite{wong2020fast}. All the data is generated from TrojAI Software Framework~\cite{karra2020trojai}. The performance is evaluated in our local machine with a 4x Nvidia Titan Xp GPU. For training data, it is available to download form ~\cite{Trojai}. A detailed configuration of each model is shown in Table~\ref{tab:dataset}. 
We need to note that for each dataset, half of the models are poisoned. Polygon triggers and Instagram filter triggers take 50\% each of all poisoned models. For the TrojAI dataset, synthetically created image data of non-real traffic signs are used for training all models. These image data include city, residential, and road. Table~\ref{tab:parameter_setup} summarizes the setting for four trigger parameters.

\begin{table}[ht]
\centering
\begin{tabular}{ll}
\Xhline{2\arrayrulewidth}
\textbf{Model architectures} & \textbf{Version} \\ \hline
Resnet  & 18, 34, 50, 101, 152 \\
Wide Resnet  & 50, 101 \\
Densenet & 121, 161, 169, 201 \\
Inception & v1 (googlenet), v3 \\
Squeezenet &  1.0, 1.1 \\
Mobilenet & mobilenet\_v2 \\
ShuffleNet & 1.0, 1.5, 2.0 \\
VGG & vgg11\_bn, vgg13\_bn, vgg16\_bn, vgg19\_bn \\
\Xhline{2\arrayrulewidth}
\end{tabular}
\caption{Round3 TrojAI dataset details} 
\label{tab:round3_dataset}
\end{table}


\begin{table}[ht]
\centering
\scalebox{0.99}{
\begin{tabular}{ll}
\Xhline{2\arrayrulewidth}
\textbf{Trigger Parameters} & \textbf{Setting} \\ \hline
Trigger Location & Random location in foreground img \\ 
Trigger Shape & Random sides and length \\ 
Trigger Color & Random RGB value \\ 
Trigger Size & Random size from 2\% to 25\% \\ 
Trigger Rotation & Random angle from 0\textdegree to 90\textdegree \\  \Xhline{2\arrayrulewidth}
\end{tabular}
}
\caption{Summary of 4 trigger parameter setup} 
\label{tab:parameter_setup}
\end{table}

Model input data should be (1 x 3 x 224 x 224). For the poisoned model, the triggers used for training is randomly generated with random sides and random length of each side. The foreground sign size is either $20\%$ or $80\%$ of the background image. As we discussed before, trigger size is a percentage of image area $2\%$ to $25\%$ uniform continuous. The dataset also provides 10 or 20 example images per class for training. Note that these images are not the exact same image used for training but built similarly. The considered model architectures are: $resnet$, $wide\_resnet$, $densenet$, $googlenet$, $inceptionv3$, $squeezenetv$, $shufflenet$ and $vggs$ with a different number of layers. The configuration of polygon trigger is shown in Table~\ref{tab:dataset}. The candidates for Instagram triggers are the same ones we consider within our approach.



\begin{figure}[ht]
    \centering
    \begin{subfigure}{0.94\columnwidth}
          \centering
          \includegraphics[width=\textwidth]{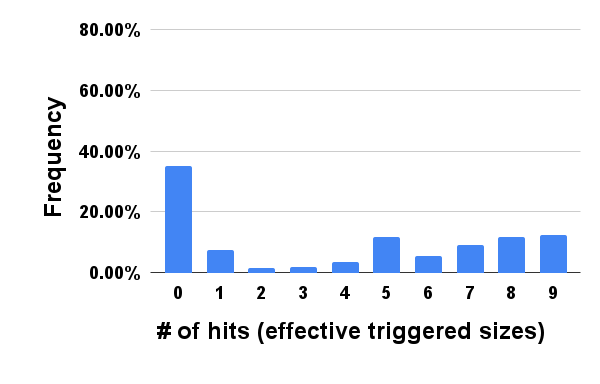}
          \caption{\label{fig:coverage_step}}
    \end{subfigure}
    \hfill
    \begin{subfigure}{0.94\columnwidth}
         \centering
         \includegraphics[width=\textwidth]{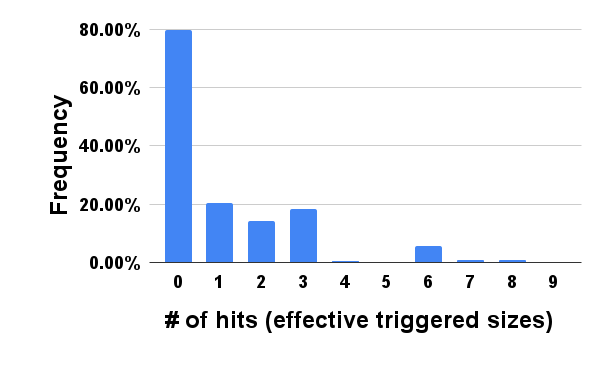}
         \caption{\label{fig:runtime_step}}
    \end{subfigure}
    \caption{Backdoored Trigger distribution. (a) poisoned model, (b) clean model.\label{fig:compare}}
\end{figure}

\begin{table}[ht]
\centering

\begin{tabular}{cccc}
\Xhline{2\arrayrulewidth}
  & \textbf{\# of models} & \begin{tabular}[c]{@{}l@{}}\textbf{\# of model} \\ \textbf{architectures}\end{tabular} & \textbf{Trigger type} \\ \hline
CIFAR10    & 10    & 1   & Polygon \& Ins. \\ 
VGGface    & 10    & 1   & Polygon \& Ins. \\ 
Train data   & 1008    & 23   & Polygon \& Ins. \\ 
Test data    & 144     & 23   & Polygon \& Ins. \\ 
Holdout data & 288     & 23   & Polygon \& Ins. \\ \Xhline{2\arrayrulewidth}
\end{tabular}
\caption{TrojAI dataset (Ins.= Instagarm)}\label{tab:dataset}

\end{table}
Figure~\ref{fig:compare} shows the trigger size distribution of the clean model and poisoned model. The x-axis represents the number of effective trigger sizes. There are total 9 trigger sizes, and the poisoned model has a higher chance to get big numbers. We can see that the distribution is somehow different between clean models and poisoned models.  A threshold of $3$ trigger sizes can offer the best classification accuracy. The probability for detecting a poisoned model with a single random color is $54.14\%$. Cumulative probability \cite{hatke1949certain} measures the chance that one single catch happens during multiple independent events. In this case, 
We leverage the cumulative probability for different numbers of independent trigger colors. The error rate is defined as a model that is falsely classified. And we set the number $4$ to balance the total error rate.

\subsection{Detection Performance}

We evaluate our method on the self-generated CIFAR, VGGFace dataset and the TrojAI round-3 train, test and holdout dataset. 
The CrossEntropyLoss$(CE-Loss)$ is defined as:
\begin{equation}
    CE-Loss = - (y\cdot \log(p) +(1 - y) \cdot \log(1-p))
\end{equation}


An ROC(Receiver Operating Characteristics) curve was originally used in signal detection theory to realize the tradeoff between the false alarmrates and hit rates~\cite{egan1975signal,green1966signal}. And this curve has been studied and applied in medical diagnosis since 1970s~\cite{metz1978basic,swets1988measuring}. AUC(Area under the ROC Curve) represents the probability that a randomly chosen negative example will have a smaller estimated probability of belonging to the positive class than a randomly chosen positive examples~\cite{hanley1982meaning,huang2005using}.
Then ROC-AUC curve is a performance measurement for the classification problems at varioud threshold settings. 

\begin{figure}[ht]
    \centering
    \includegraphics[width=0.48\textwidth]{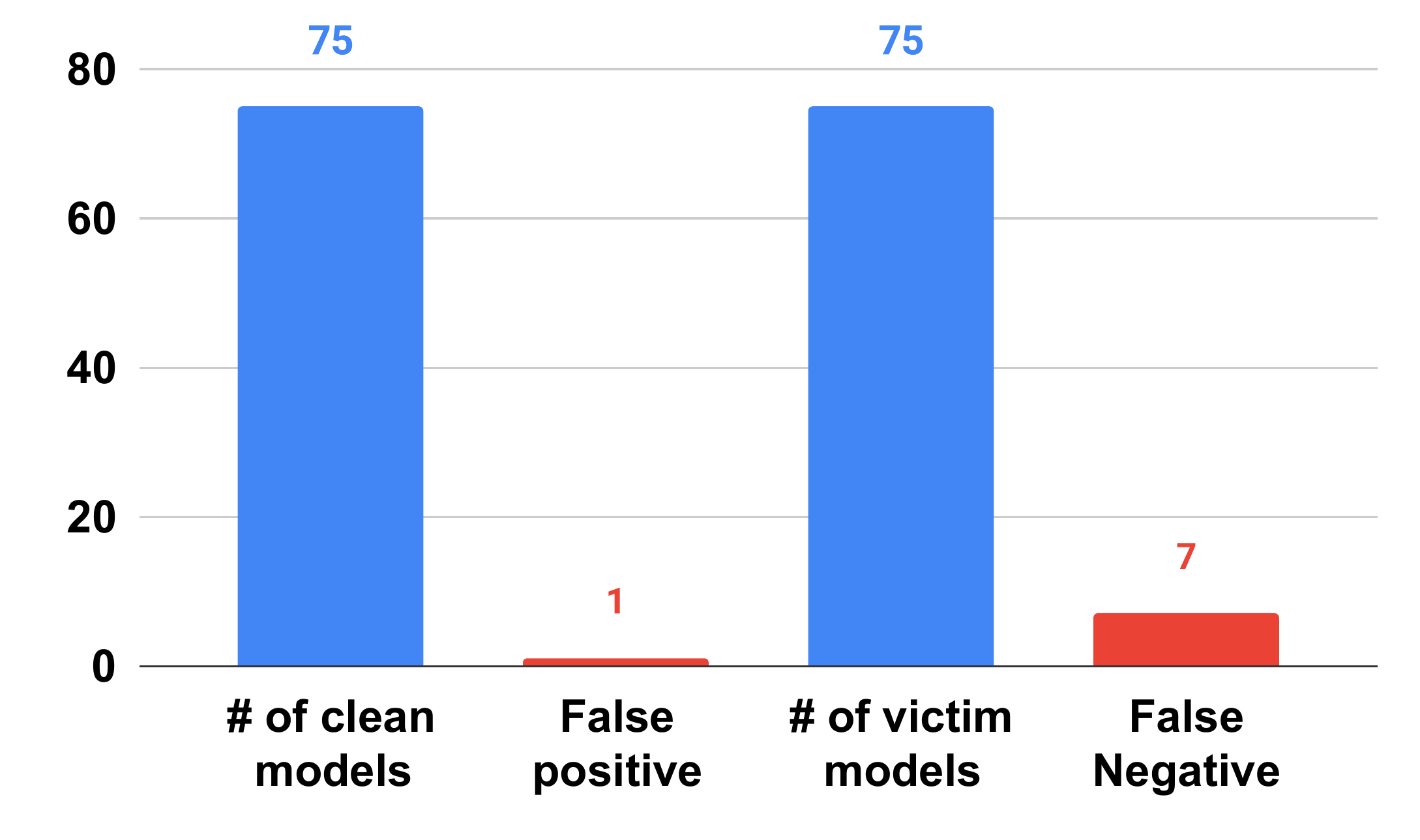} 
    \caption{Evaluation performance for models with polygon triggers and clean models.}
    \label{fig:poly_only}
\end{figure}

We first test our performance for models with Polygon triggers and models with Instagram filter triggers individually. Figure~\ref{fig:poly_only} shows the result for 75 models with polygon triggers and 75 clean models from TrojAI train dataset. The output ROC-AUC is 0.95. CE-loss is 0.222.

In Figure~\ref{fig:poly_only}, we can see that there is 1 false positive and few false negatives, those false negatives can be resolved by increasing the rounds of random color but we make a runtime trade-off to achieve acceptable results.

\begin{figure}[ht]
    \centering
    \includegraphics[width=0.48\textwidth]{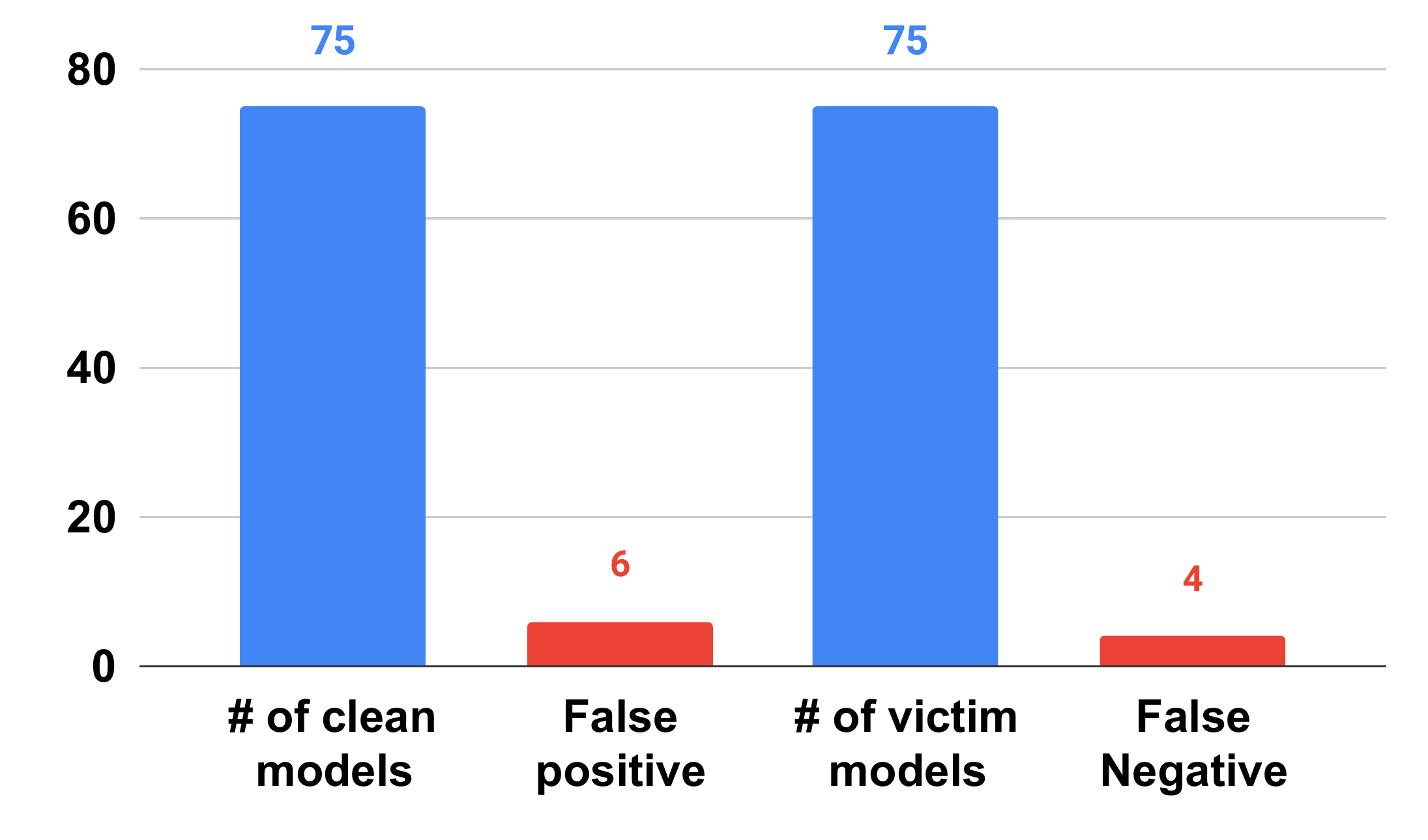} 
    \caption{Evaluation performance for Instagram-only models.}
    \label{fig:Ins_only}
\end{figure}

Figure~\ref{fig:Ins_only} shows the result for 75 models with polygon triggers and 75 clean models from TrojAI train dataset. We get 0.933 ROC-AUC and CE-loss is 0.2513.

In Figure~\ref{fig:Ins_only}, we can see that there is few false positives, we believe they are "natural" Trojan inside of clean models. In other words, some clean models introduce backdoor during the training step even without adversarial data.

From the figure, we can see that there is 1 false positive and few false negatives, those false negatives can be resolved by increasing the rounds of random color but we make a runtime trade-off to achieve acceptable results.

Table~\ref{tab:train_eva_data} shows the result of both train data and test data. For Id-00000000-099 models, they consist of 50 victim models and 25 models have polygon triggers and 25 models have Instagram filter triggers, the rest 50 models are clean models. The same case applies to Id-00000100-199 models and test models. We can observe that our proposed method produces consistent performance without overfitting. 




We further evaluate our method for holdout data and compare it with two state-of-the-art methods. Table~\ref{tab:abs_comparasion} shows the evaluation results. Note that since NC \cite{wang2019neural} uses the "minimal" trigger idea in a fixed location and reverse engineering to produce $N$ potential "triggers", it does not work for our model settings at all. Besides, our results show that false predictions are primarily false positives and half of the false positives are detected as victim models while the other half are detected as Ins model. This is a fascinating situation and we need to deal with the model individually. Therefore We have been trying different methods to minimize these false positives. The method is very tricky because the false positives have almost identical responses compared to the Trojaned models. We suspect these models are "fake" benign models and we still need to dig into it to find the hidden difference between Trojaned models and "fake" benign models. 


We compare our result with a state-of-the-art method ABS\cite{liu2019abs} and 95\% confidence interval for our cross-entropy loss (CE-95\% CI) is introduced for comparison. Table~\ref{tab:abs_comparasion} shows the result and we can see that our method is better than ABS in terms of the CrossEntropyLoss, which is the essential score to evaluate a method. Note that our method is a black-box based method and ABS is a white-box based method. To the best of our knowledge, our approach is the first black-box method that can address some complex settings. 

\begin{table}[ht]
\centering
\scalebox{1}{
\begin{tabular}{llll}
\hline
\textbf{Dataset Models} & \multicolumn{1}{l}{\textbf{CE-Loss}} & \multicolumn{1}{l}{\textbf{ROC-AUC}} & \multicolumn{1}{l}{\textbf{Runtime(s)}} \\ \hline
CIFAR10 & 0.00 & 1.00 & 47.4  \\ \hline
VGGFace & 0.00 & 1.00 & 119.6  \\ \hline
Train Models (TrojAI dataset)  & \multicolumn{1}{l}{} & \multicolumn{1}{l}{} & \multicolumn{1}{l}{} \\ \hline
Id-00000000-099 & 0.2367 & 0.94 & 234000  \\ \hline
Id-00000100-199 & 0.1928 & 0.96  & 302727 \\ \hline
Test Models (TrojAI dataset) & \multicolumn{1}{l}{} & \multicolumn{1}{l}{} & \multicolumn{1}{l}{} \\ \hline
288 Test models & 0.3184 & 0.9027 & 372120\\ \hline
\end{tabular}
}
\caption{\label{tab:train_eva_data}Evaluation on different data.}
\end{table}


\begin{table}[ht!]
\centering
\begin{tabular}{lllll}
\hline
\textbf{Methods} & \textbf{Models} & \multicolumn{1}{l}{\textbf{CE-Loss}} &\textbf{CE-95\% CI}& \multicolumn{1}{l}{\textbf{ROC-AUC}} \\ \hline
Our approach & 288 models & \textbf{0.3184} & 0.0754 & 0.9027 \\ 
ABS\cite{liu2019abs} & 288 models & 0.3552 & 0.0717 &  0.91037 \\ 
NC\cite{wang2019neural} & 288 models & $\diagdown$ & $\diagdown$ &  $\diagdown$ \\ \hline
\end{tabular}
\caption{\label{tab:abs_comparasion} Comparison with the-state-of-the-art method.}
\end{table}





Even though our method enables very high performance, it still has some drawbacks. For example, there is a very small $0.02$ difference of ROC-AUC because of the randomness of color initialization, which will be improved in future work.

%% file: 6_discussion.tex
\section{Discussion}
In this section, we present the sensitivity analysis for our approach and plot different figures for different kind of parameters. Then we discuss the complexity of our approach. Finally, we list the trade-offs in our method.

\subsection{Sensitivity analysis}

\subsubsection{Rounds vs performance}
\begin{figure}[ht!]
    \centering
    \includegraphics[width=0.48\textwidth]{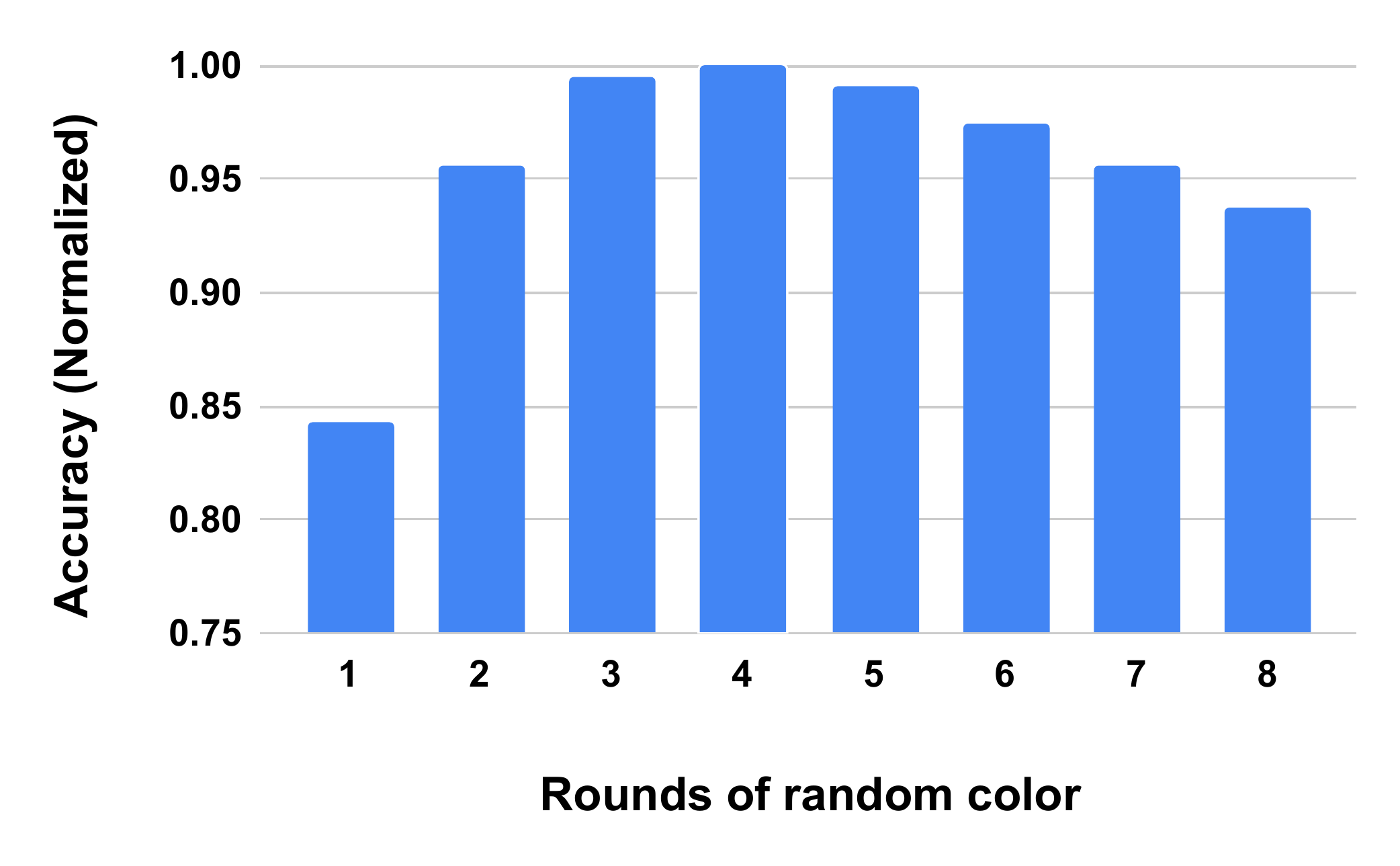} 
    \caption{Success rate for different rounds of experiment.}
    \label{fig:success_rate}
\end{figure}

Figure~\ref{fig:success_rate} shows the detection accuracy for 50 models with polygon triggers and 50 clean models from the training dataset. We fix the trigger location to the center of the image and trigger step size is $0.02$. The accuracy goes up until $rounds = 4$ and then goes down due to the accumulative false positives. We choose $rounds = 4$ for our experiment. Since there is a trade-off between the number of random color rounds and runtime, we should consider this situation in practical scenario. Note that the detection accuracy for models with polygon triggers is lower than Instagram filter models, which offers almost 100\% accuracy. 

\subsubsection{Rounds vs runtime}

\begin{figure}[ht!]
    \centering
    \includegraphics[width=0.48\textwidth]{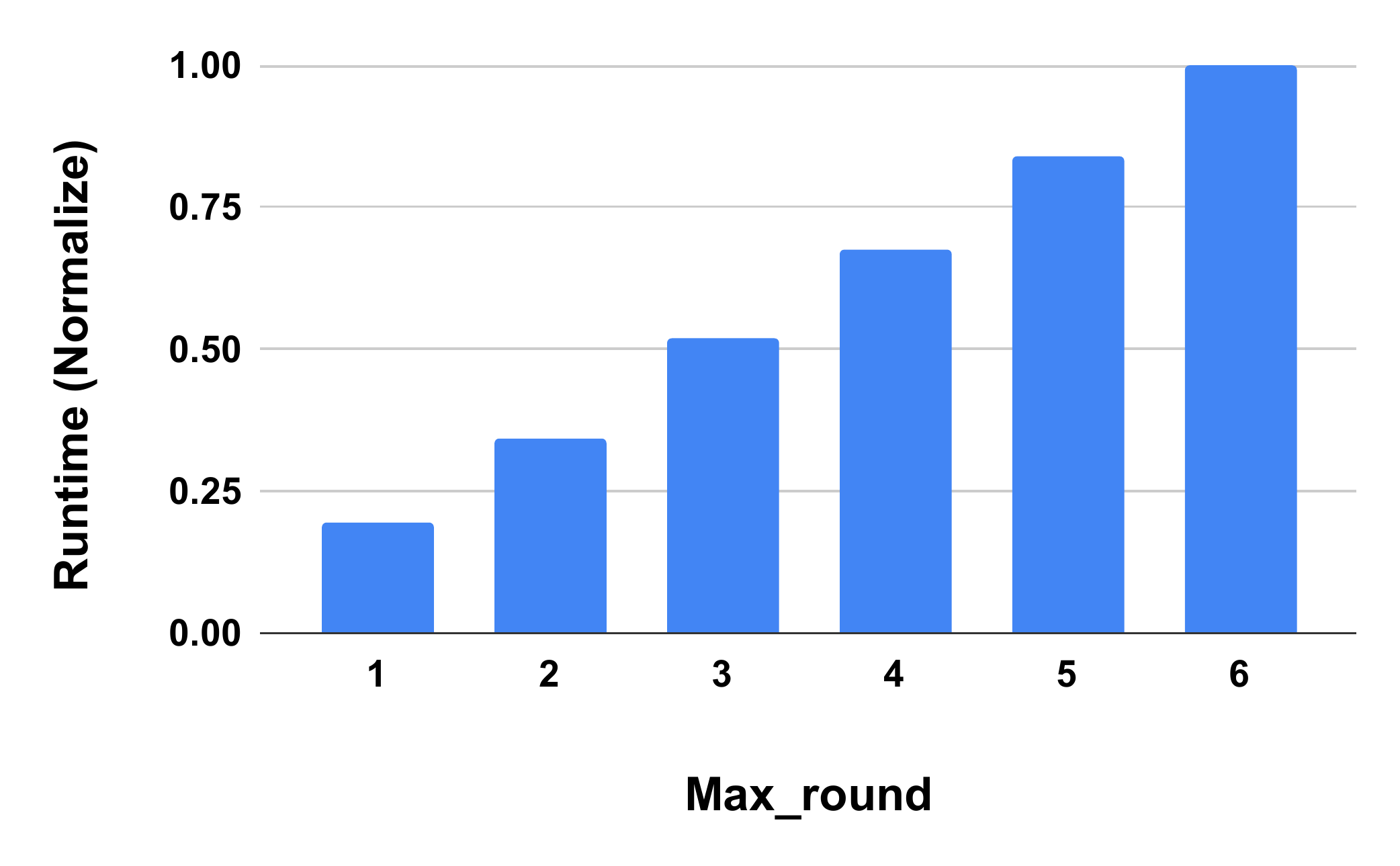} 
    \caption{rounds vs runtime.}
    \label{fig:rounds_runtime}
\end{figure}

Figure~\ref{fig:rounds_runtime} shows the runtime for different rounds of random sample colors. We can say from the figure that as rounds goes up, the runtime increases linearly, which is acceptable for practical case that the model can be verified with in several minutes before deployment. We believe that the time around 30 minutes is an acceptable time limit in practical case.

\subsubsection{Number of trigger areas vs performance}
\begin{figure}[ht]
    \centering
    \includegraphics[width=0.48\textwidth]{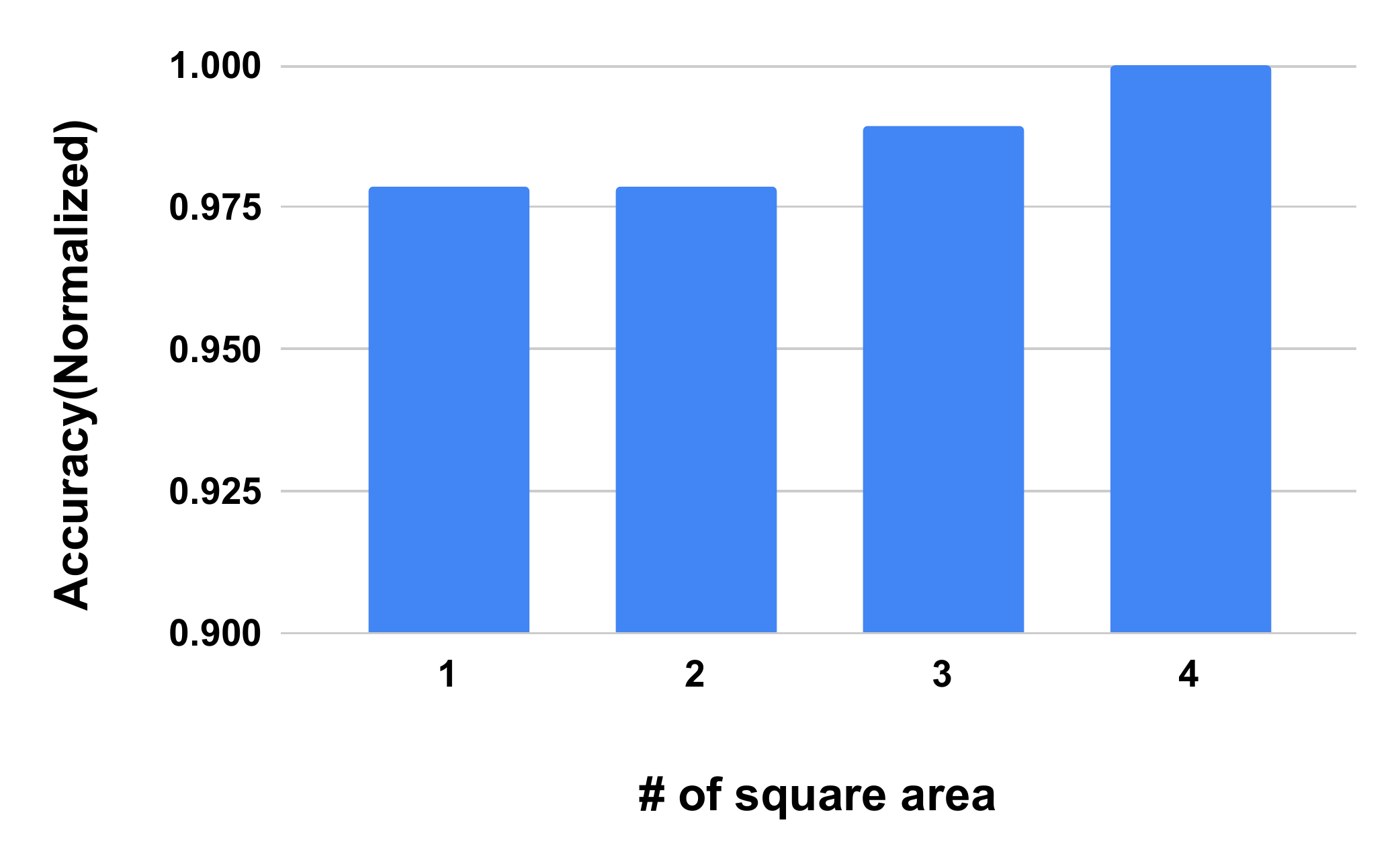} 
    \caption{Number of trigger areas vs performance.}
    \label{fig:areas_performance}
\end{figure}

Figure~\ref{fig:areas_performance} show the plot between performance and number of trigger areas. As we mentioned before, some trigger areas does not play am important role in classifying the poisoned models. Note that we pick the best areas in 4 trigger areas. 1 or 2 trigger areas does not change much for the performance while 3 trigger areas slightly improve the performance of our proposed methods. 4 trigger areas offers best performance but it takes 3x more times compared with the 1 trigger area. We suggest use 1 trigger areas for quick detection and 3 trigger areas for a better detection. If the runtime is not a big issue, 4 trigger areas would offer the best performance.

\subsubsection{Step-size for trigger size vs performance}

\begin{figure}[ht!]
    \centering
    \includegraphics[width=0.48\textwidth]{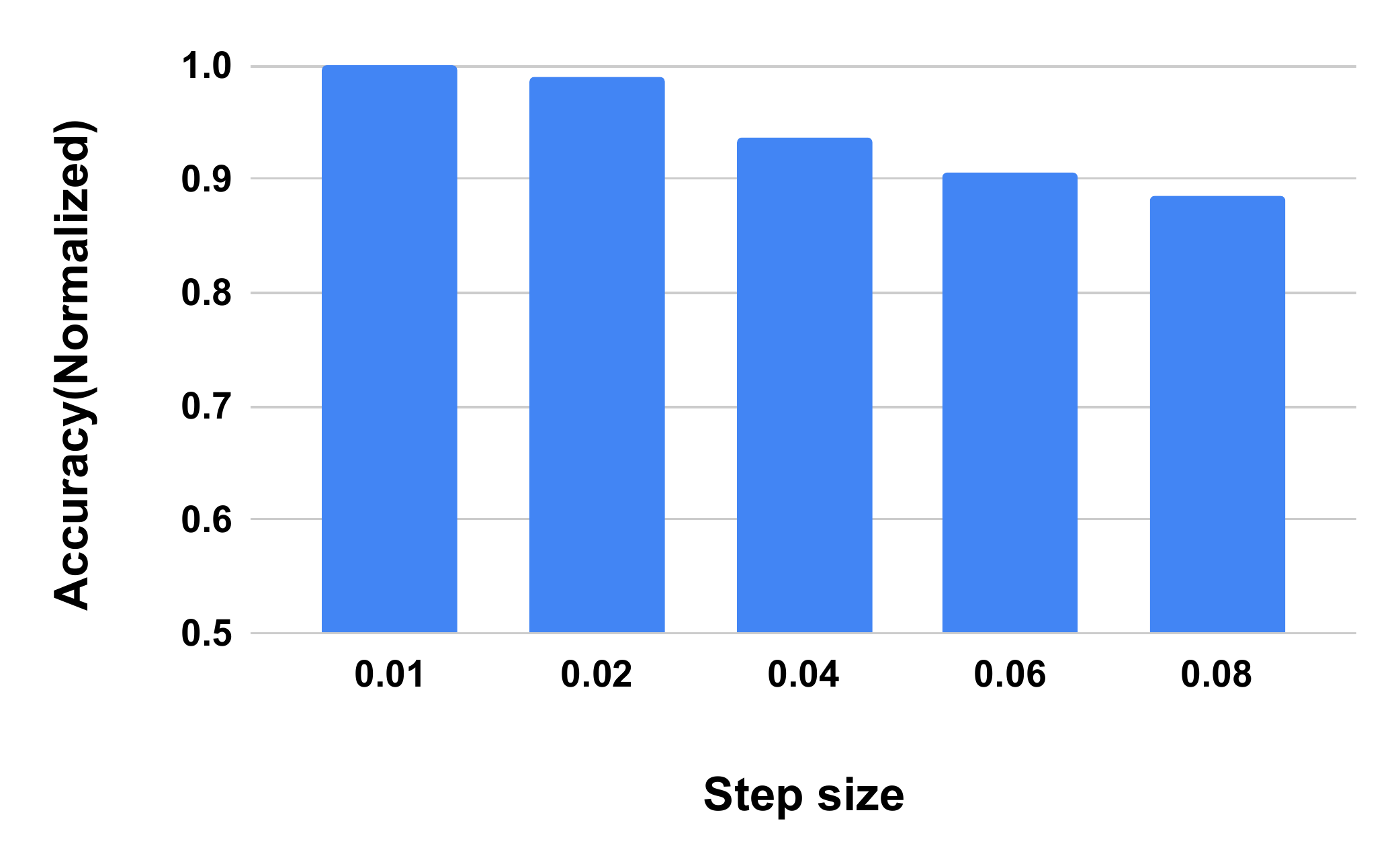} 
    \caption{Different step sizes vs performance.}
    \label{fig:stepsize_performance}
\end{figure}

Figure~\ref{fig:stepsize_performance} shows the trade-off between stepsize of trigger sizes and performance. We can see from the figure that as stepsize increases, the performance decrease while the runtime reduces.

%% file: 7_conclusion.tex
\section{Conclusion} \label{sec_conclusion}

In this paper, we proposed the first trigger approximation based black-box  Trojan detection framework that enables fast and scalable search of  the trigger in the input space. Moreover, our approach also detects Trojans embedded in the feature space where certain filter transformations are used to activate the Trojan. Empirical results show that our approach achieves superior performance than other state-of-the-art methods.

Our future direction will cover the following parts. a)Improving the consistency of output accuracy and runtime performance. b) Applying our approach to more tasks such as NLP. c) Investigating natural Trojans in clean models.